# Long-Term Prospects: Mitigation of Supernova and Gamma-Ray Burst Threat to Intelligent Beings

Milan M. Ćirković[1]
Branislav Vukotić
*Astronomical Observatory of Belgrade, Volgina 7,*
*11000 Belgrade, Serbia*

**Abstract.** We consider global catastrophic risks due to cosmic explosions (supernovae, magnetars and gamma-ray bursts) and possible mitigation strategies by humans and other hypothetical intelligent beings. While by their very nature these events are so huge to daunt conventional thinking on mitigation and response, we wish to argue that advanced technological civilizations would be able to develop efficient responses in the domain of astroengineering within their home planetary systems. In particular, we suggest that construction of shielding swarms of small objects/particles confined by electromagnetic fields could be one way of mitigating the risk of cosmic explosions and corresponding ionizing radiation surges. Such feats of astroengineering could, in principle, be detectable from afar by advanced Dysonian SETI searches.

## 1. Introduction: risks from SNe and GRBs

### 1.1. Cosmic explosions as existential risks

The concept of existential risk (and somewhat wider and vaguer category of global catastrophic risks) has been recently brought into focus of much discussion, some of which is highly relevant for astrobiology, e.g. [1-4]. Any biosphere in the Galaxy is necessarily exposed to a wide spectrum of natural hazards; if the biosphere includes intelligent beings and their civilizations, the list of threats also includes additional items – generalized analogs of what nowadays on Earth are called anthropogenic risks. The totality of all threats could, in principle, be described by a total risk function which would be a function of spatial location of any habitat in the Galaxy and epoch in the Galactic history. While the construction of such a risk function remains as a task for astrobiologists of the future, it is of paramount importance to study all hazards facing potential biospheres throughout the Galaxy for both theoretical and practical reasons.

---

[1] Also of the Future of Humanity Institute, Faculty of Philosophy, University of Oxford, Suite 8, Littlegate House, 16/17 St Ebbe's Street, Oxford, OX1 1PT, UK. E-mail: mcirkovic@aob.rs.

One of the global catastrophic risks from nature, with possibility to ascend to the level of existential risk, is the one of *cosmic explosions*, notably close supernovae (SNe) and γ-ray bursts (GRBs). Both are natural consequences of stellar evolution. Supernovae (known as an observational phenomena since ancient Chinese astronomers; e.g., [5] are either thermonuclear explosions of white dwarfs in close binary stellar systems triggered by the post-Main sequence evolution of the companion star; or terminal explosions of very massive stars (for both types there exists a massive literature; see, e.g., reviews in [6,7]. Transient sources visible in the Milky Way in AD1006, AD1054, AD1572, and AD1604, and in the M31 galaxy in AD1885 were all SNe. GRBs are much less understood, but the extensive research in the last half century suggests that they also come in two types: shorter ones associated with the terminal merger of binary neutron stars and longer ones representing gamma-ray signatures of *hypernovae*, or the terminal explosion of supermassive stars, analogous to Eta Carina in the Milky Way [8-10]. Due to this latter mechanism, there is a form of continuity over the spectrum of terminal stellar explosions, e.g. [11], justifying their unified treatment from the standpoint of risk analysis. The main reason why the subject has not been covered extensively – even when compared with other huge and rare natural hazards like asteroidal/cometary impacts or supervolcanic eruptions – has been the extreme rarity of such events, lack of proof for historical occurrence even on geological timescales, the lack of scaling modes, and uncertainty about the ecological effects of such cosmic explosions. Very recent discoveries of various types of previously unknown ultra-luminous supernovae, such as the controversial case of ASASSN-15lh, presented by Dong et al. [12], with peak bolometric luminosity of about $2.2 \times 10^{45}$ erg s$^{-1}$ (about $5.7 \times 10^{11}$ L☉), indicate a whole new family of such astrophysical sources [13]. These new findings can only strengthen the case for the astrobiological importance of such cosmic explosions.

## 1.2. Explosion risk in the Galactic Habitable Zone

In some of our previous work [14-19], we have considered the possibility that it is exactly this type of catastrophe which could serve as a global regulation mechanism, preventing the emergence of life in distant past of the Milky Way. Following upon the hypothesis of Annis [20], this might serve as an explanation of Fermi's paradox, namely the absence of *much older* advanced technological civilizations and their traces and manifestations in the Galaxy. If the regulation mechanism secularly evolves toward increased habitability, the emergence of such societies characterizes only present and future, not our past. Since cosmic explosion rates have generally been exponentially declining over cosmic time, as testified by cosmological observations, as well as models of star formation, e.g., Yungelson and Livio [21], they present a viable candidate for such global regulation mechanism. One of the ingredients in this picture is the assumption that truly advanced societies will be able to utilize their engineering capacities in order to make themselves safe against *exactly the type of threat underlying the regulation mechanism*. If those threats are cosmic explosions, then we clearly need at least a vague argument that advanced technological civilizations could become plausibly immune against the adverse effects of such explosions. In other words, the regulation mechanism needs a cut-off or threshold. This has not been covered in the literature thus far, neither in SETI-related discourse, nor in the domain of risk analysis and mitigation. The present study aims at filling that gap.

(An exception in this respect is the study of Leggett [22] which, in a broad and instructive overview of existential risk threats, takes the risk of Galactic GRBs very seriously and advise that we create deep underground shelters for a representative cross-section of lifeforms. At the end of the relevant section, he laconically adds: "Better yet, with future technology, an Earth-protecting space shield may be feasible within, say, 10 000 years, other risks permitting." (p. 789) Although on the right track, we find this timeline unnecessarily pessimistic.)

The comparison of these three types of global catastrophic risks from nature – impacts, supervolcanoes, and cosmic explosions – is instructive in several ways. All three, we now know, operate on long timescales, characteristic for both astrophysics and geosciences. While the impact hazard has been acknowledged only very recently – essentially only after the seminal Alvarez et al. paper [23] and the ensuing shift towards neocatastrophism [24-26] – the literature on both prediction and mitigation of impact threat has already grown to sizeable proportions. This has been helped by the fact that impact risks *scales* down to local events, analogous to the 1908 Tunguska explosion [27], and even further down to analogs of the 2013 Chelyabinsk air burst. Supervolcanism has been only relatively recently recognized as a major natural risk of astrobiological importance [28,29]. It scales down even better to well-known historical volcanic explosions, such as Mount Tambora in 1815, Krakatoa in 1883, or Mount St. Helens in 1980.

In sharp contrast to the impact and supervolcanism risks, the cosmic explosion risks does not scale down to more frequent and tame events, easily observable on human life timescales and in comfortable vicinity. Physics of SNe/GRBs does not allow for such a possibility – and massive stars are anyway extremely rare even within the Galactic disk. The best analogue to smaller events are magnetar explosions which have only very recently been recognized, motivated by the 27 December 2004 flare of the soft γ-repeater SGR 1806-20 [30,31]. Considering the fact that the source of the burst has been located at distance kpc, those minor repeating explosions seem to have more potential to create astrobiological and ecological effects throughout the Galaxy. Since they are clearly orders of magnitude more frequent than terminal explosions, the conclusion about importance of explosive radiation events for astrobiology can only be strengthened.

There is a consensus of researchers that sufficiently close cosmic explosions can have disastrous effects on any biosphere, including the terrestrial one; the controversy is just how often is an average biosphere is subjected to such a stress, at which distances cosmic explosions are fatal, and what are the net effects of such a stress in non-fatal cases. The potentially catastrophic impact of nearby Galactic supernovae on life on Earth was noted many decades ago, e.g. [32-35], and even some particular extinctions were ascribed to causal mechanisms triggered by cosmic explosions [36-39]. This scales down from the Gyr and Myr timescales down to last ~300 Kyr [40,41], and even as late as the Younger Drias extinction/climate change event ~12,000 years ago [42-43]. However, it is only in the last two decades that serious models of the threat have been developed, when the advances in both stellar astrophysics, observational cosmology and astrobiology have enabled better understanding of radiation intensity and composition [44,45] as well as ecological and physiological impact of ionizing radiation and other consequences of cosmic explosions on biospheres [39,46-55]. All this research activity clearly testifies that the awareness of the astrobiological significance of cosmic explosions has surged dramatically in recent years and that the topic is not any more dismissed or stigmatized as science-fiction.

To illustrate the importance and severity of the threat from energetic explosions we analyse the data from the simulation presented in Vukotić et al. [56]. They analysed the snapshots from the *Gadget2* smoothed-particle hydrodynamical (SPH) *N*-body simulation of an isolated spiral galaxy to estimate the probability of finding Earth-similar habitats and determine the extent of the Galactic Habitable Zone in a generic Milky Way-like galaxy. This is one of the first instances of application of cosmological structure-formation simulations in astrobiology. For the purpose of this work, we apply similar but somewhat simpler procedure. We use the calculated surface density of the star formation rate (SFR) for the galactic disk from [56] and investigate how much time particles spend in the regions of particular SFR. In Figure 1, we plot on the Y-axis the number of the candidate habitable particles (CHP) with galactocentric radii between 3 and 20 kpc, that have spent more than a particular fraction (colour-coded) of their life time in the areas of the galactic disk with a SFR that is higher than indicated on the X-axis. If a CHP moves below the galactocentric radius of 3 kpc we consider it to reside in the bulge of the galaxy, and therefore in the area of SFR higher than indicated on the X-axis. Contrariwise, if the galactocentric radius is higher than 20 kpc, then that particular CHP is considered to be in the area with SFR smaller than the corresponding X-axis value. The lines on the plot are for the CHPs that have spent more than 80, 60, 40 and 20 % of their lifetime in the areas with a SFR higher than indicated on the X-axis. The black solid line plots the total number of CHPs, regardless of the mentioned life time fraction. The dashed black line is the surface SFR density from the Solar neighborhood, 0.005 Solar masses year$^{-1}$ kpc$^{-2}$, as adopted in [56]. We plot for the time instances at 5, 7 and 10 Gyr of the simulation time.

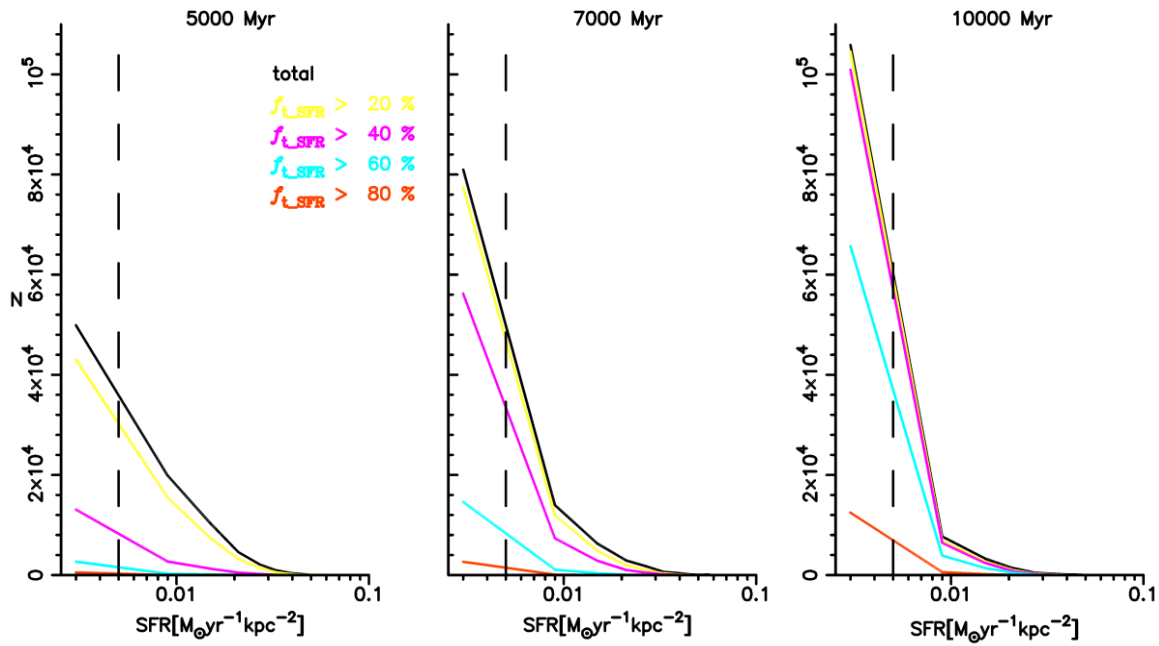

**Figure 1.** Simulated candidate habitable particles in regions of different star-formation rate per fraction of their lifetimes. Panels represent different epochs from the simulated "galaxy formation" epoch.

Since high SFR regions correspond to regions where the threat of radiation events is high, e.g., spiral arms in *present-day* spirals as well as analogous regions at high redshift [57,58], we wish to minimize the fraction of time spent in those regions if we seek *continuously* habitable stars and planetary systems. We notice that at later epochs in the galactic history (central and right-hand plot) the behaviour of yellow and pink lines is not very different from the black line, indicating that the risk of being in the vicinity of a supernova or a GRB is large for the large fraction of candidate habitable particles. Coupled with the fact that it is reasonable to assume – Earth's single case notwithstanding – that evolving intelligent beings, not to mention technological civilizations, requires timescales on the order of several Gyr, during which a single close explosion could cut or derail the evolutionary chain of events leading to that outcome, this indicates that once a civilization emerges, it is only rational to seriously consider risk from such events and possibilities of its mitigation. While the resolution of the simulation of Vukotić et al. [56] is still insufficient for conclusions about number of individual stars and planetary systems, it is still indicative and motivating for further work in the area. For the present purposes, we note that no part of any spiral galaxy can be considered safe from cosmic explosions in the long run. And as the timeframe considered by an intelligent species grows longer, more relevant becomes the issue of mitigation. It is reasonable to hope that near-future simulations of habitability will offer more complete and precise account of the amount of risk faced by different parts of the Galactic Habitable Zone.

## 1.3. Preconditions for mitigation

An absolutely essential precondition for *any* type of mitigation effort is sufficiently reliable predictive power: we need to know precisely *when* the explosion will go off, *how strong* will it be, and *how isotropic* (or not) will the emission of all relevant products be. The advances in stellar astrophysics have been tremendous in recent decades and while we are now in position to build more or less detailed models of Type II/Ic supernovae (and corresponding longer GRBs), there is still a long way to go until we are able to put a specific timescale to each individual progenitor star; this is obvious even in the case of so well-studied progenitor as η Car. Timing is even more important here considering how brief these sources are in comparison to almost everything else in astrophysics – and this immediately tells us that the actual determination of timing for a cosmic explosion is likely to be influenced to an unusual degree by factors usually neglected in astrophysical context. In contrast to the other two vital pieces of information, where a simple Boolean output might be sufficient for practical purposes (*is the explosion strong enough, for given distance, enough to warrant mitigation measures?* YES/NO; *are we located within the solid angle in which most of the energy is emitted?* YES/NO), there is no shortcut in the need to predict – or retrodict – the exact timing of the event. This is clearly beyond the scope of even the best numerical simulations of today.

However, there is no reason to doubt that *barring other defeaters* any future human civilization – or an extraterrestrial civilization at the analogous development stage – will be able to understand the physics of SNe and GRBs to much higher degree of precision. Whether it will enable *sufficiently* exact predictions of their explosion times, energies, spectra, cosmic-ray acceleration power, etc., remains to be seen, but there is no obstacle of principle in this matter. While prediction of weather in its local detail is still notoriously uncertain, the trajectory and timing of hurricanes, cyclones and other storm systems storms is today routinely predicted, often enough in advance for efficient mitigation measures to be deployed. History of science offers many examples of the increase of reliability and accuracy of predictions in various other areas, from eclipses to neutrino pulse from supernova SN 1987A. Even in the areas where predictions have not built a good track record so far (e.g., earthquakes, volcanic eruptions, economic crises), we are gradually focusing on the main obstacles to further progress and the development of massive numerical simulations did much to understand the related problems much better.

While the present-day humanity could hardly do anything useful in the case of a nearby cosmic explosion (except, perhaps, preserving a bunch of time-capsules for hypothetical posterity or extraterrestrial visitors in distant future), we shall, in further text, consider mostly a humanity-like civilization which is moderately more advanced in terms of space technology. This civilization – which we shall dub **humanity+** – could undoubtedly emerge on several decades, up to a century future timeframe, barring global war, deep economic crises or other catastrophic defeaters. Since the timescale for close SNe/GRBs is much longer than that, probability of one going off in the next century is negligible (and in any case much smaller than the probability of other global catastrophic risks), and the probability density stays the same over such an interval. We suppose that humanity+ is still mostly located on Earth in terms of population and industrial resources, while utilization of resources of the Solar System bodies, notably Moon, Mars, asteroids, comets, etc. is under way. Also, no revolutionary technological breakthroughs/miracles are assumed – no "warp drives", "photon rockets", or "zero-point energy engines". In other words, **humanity+** is a placeholder for a small civilization making its first serious steps into interplanetary space.

The other case of possible interest – if only to serve as a contrast as to what is realistically possible in the near future – is a case of very advanced technological civilization, capable of interstellar flight and astroengineering at interstellar scales. We can call it **humanity++** and can speculate whether it will need centuries or millennia or even longer to achieve (barring defeaters); there is no real sense in speculating whether such a civilization would be of biological or possibly superintelligent AI nature or about its other properties.[2] It is enough to consider its capacities of changing its (astro)physical environment on large scales of space, time, and energy. In the simplest and most conventional rendering, **humanity++** could build a Dyson shell around the Sun or some other convenient star in the Milky Way, engage in stellar uplifting or rejuvenation, construct habitable shellworlds or ringworlds, and in general do any number of astroengineering feats which are still "wild cards" for us at present [61-70]. Some of them would, by its very nature, be impervious to most or all detrimental effects of cosmic explosions – for instance, Dyson shells, shellworlds or supermundane planets will be resistant to electromagnetic ionizing radiation and all but the highest energy cosmic rays.

In the rest of this cursory study, we shall consider possible approaches to the mitigation of the cosmic explosion threat. The treatment is very preliminary, since there is so much uncertainty about both the Galactic distribution and local ecological effects of these events, and the sample is either too small to allow for any statistical analysis (historical supernovae close to the Solar system) or too dependent on the unknown properties of their environment (extragalactic supernovae/GRBs). Even more uncertainty surrounds means and capacities of advanced technological civilizations, future human or extraterrestrial. Therefore, the essential purpose of this study is to roughly outline the problem itself and suggest one – quite crude – approach to resolving it. It is important to emphasize from the outset that while details of the interaction of cosmic explosions with biospheres are still largely unknown, they are of minor importance for the central goal of this paper. As Ludwig Boltzmann [71] famously said: "It may be objected that the above is nothing more than a series of imperfectly proved hypotheses. But granting its improbability, it suffices that this explanation is not impossible. For then I have shown that the problem is not insoluble, *and nature will have found a better solution than mine*." [present authors' emphasis] We would add only that "nature" here should be expanded to encompass actions of advanced technological civilizations – which may or may not be recognizable as such [72].

---

[2] A more general classification is, of course, due to Kardashev [59]. Our categories **humanity+** and **humanity++** can be regarded as a simplified cases of a range of Kardashev's Types as applied to future humanity. Thus **humanity+** might correspond to Kardashev Type 1.x, while **humanity++** is equivalent to Type 2 and above. Human civilization was about Kardashev's Type 0.72 in 2012 [60].

## 2. Mitigation approaches

Suppose that at some point in future astrophysical data indicate that a classical supernova at the distance of several parsecs or a GRB/hypernova at the distance of several kiloparsecs will occur at a given location in a precisely given timeframe. Further, suppose that contemporary human or posthuman civilization has resources and willingness to undertake the mitigation task. In such a case, three logically possible approaches are possible:

(A) **Mitigation *in situ*:** measures undertaken at the source, preventing or containing the explosion.
(B) **Local mitigation:** measures undertaken on Earth or in its immediate vicinity.
(C) **Intermediate mitigation:** measures undertaken at some point between Earth and the source.

Option (A) is clearly infeasible for humanity and **humanity+**. It would need not only routine interstellar travel, but also astroengineering capabilities clearly belonging to the **humanity++** domain.

One such possible feat of astroengineering has been described in the fictional context by Alastair Reynolds in his *House of Suns* as creating a solid Dyson shell around the exploding star, literally *containing* the blast, Reynolds [73]. While this could be a challenge even for **humanity++**, especially in terms of material science, one should keep in mind that Reynolds' story takes place *billions* of years into the Galactic future in the context of what is essentially Kardashev's Type 3 civilization.

Less extravagant ideas include using of stellar uplifting [74], possibly along the lines suggested for extension of stellar lifetimes [65,66]. Decrease in stellar mass would lead to reorganization of stellar structure on short timescales resulting in increased lifetime which roughly scales as $\tau \propto M^{-3}$. Obviously, this requires a prolonged action in advance of the predicted terminal explosion – and since the most massive stars, which are progenitors of Type II/Ic supernovae and GRBs (at least the longer ones), anyway live very short lives ($10^6$-$10^7$ years), contingency planning would need to start very early in the "risky star's" life and include extremely intensive removal of matter. While similar ideas are exciting to contemplate – and even model quantitatively and look for elsewhere in the Galaxy as part of the Dysonian SETI activities – they are clearly very remote in the case of humanity, thus being subject to all kinds of uncertainties characterizing long-term predictions.

The option (B) is essentially analogous to what has been suggested for lunar settlements in case of Solar flares and eruptions: go below ground into the appropriate shelter for the duration. Solar weather forecasting enables prediction of extreme events on our star which create increased cosmic ray fluences usually lasting several days [75]. However, this is hardly practical in the case of SN/GRB threat, for several reasons. While the duration of gamma-ray emissions is rather short (< 100s), the pulse of accelerated cosmic rays is likely to be much longer and to extend over months or even years in the aftermath of the burst. Not only is underground sheltering of a large population for such prolonged time impractical, it would have to occur at prohibitively large depths, since cosmic-ray jets are likely to penetrate up to ~3 km within the crust of Earth or a similar Earth-like planet before dropping to less than 1% of the incoming flux, e.g. [76,77]. Even more importantly, the danger to Earth concerns to a large degree the changes induced in the atmosphere, notably the creation of NOx and subsequent destruction of the ozone layer with catastrophic ecological consequences. A planetary atmosphere cannot be hidden within a shelter – at least not globally. Finally, by the time **humanity+** emerges, there will presumably be important installations and assets to protect outside the home planet, but still in the domicile planetary system and within the same solid angle where the bulk of radiation is emitted. Those assets (e.g., orbital stations, asteroid mining facilities, important interplanetary communications hubs, even O'Neill habitats) might not be capable of active motion and thus using planets as natural shielding might not be a practical option.

This leaves us with option (C), intermediate mitigation. Obviously, this way of mitigating the cosmic explosions risk implies erecting some sort of shielding analogous to the protection of spacecraft or orbital stations against "normal" solar radiation and cosmic rays. It is to this most sensible shielding option that we turn in the rest of this paper.

## 3. Shielding planets?

If the best strategy for mitigation of cosmic explosion risk is shielding various valuable targets, including inhabited planets, several pathways are possible from there. The first issue one confronts is the choice of material for the shield. As far as high-energy γ-rays are concerned, the Table 1 summarizes the linear attenuation coefficients for some materials discussed in the literature at 0.1 MeV and 1 MeV. The bulk of radiation from the cosmic explosions is emitted in this interval of photon energies. The transmitted radiation intensity at fixed energy is given as:

$$I(E) = I_0(E) \exp\left[-\mu(E) x\right], \qquad (1)$$

where *x* is the path length within material, and *μ(E)* is the linear attenuation coefficient at energy *E*, given for some materials in Table 1.

| E (MeV) | granite | lead | ice | Wood | carbon (graphite) | air | copper |
|---|---|---|---|---|---|---|---|
| **0.1** | 0.4 | 5.55 | 0.171 | 0.085 | 0.3388 | 0.0002 | 3.9 |
| **1** | 0.12 | 0.77 | 0.071 | 0.032 | 0.1399 | 0.000085 | 0.51 |

**Table 1.** Linear attenuation coefficient for some materials in cm$^{-1}$, as measured by NIST [78]. Stony asteroids are similar in their attenuation properties to granite, and an "average" ice has been used.

Perhaps the most natural building material for structures in the outer Solar system is ice [79]. Ice is quite abundant, especially in satellite/ring systems of gaseous and ice giants [80] and in the Kuiper Belt where it has been detected in several individual objects, e.g. [81]; most of these objects have colours and spectra consistent with icy composition similar to the nuclei of well-studied comets [82]. Amorphous high density ice phase (I$_a$h) has density of about 1.1 g cm$^{-3}$ at very low temperatures, pp. 372-374 in [79], not taking porosity into account.

Since the Kuiper Belt is a low-gravity, low-temperature environment, more convenient for engineering purposes, we suggest that it is the most convenient site for construction of shielding swarms. The radius of the parent body which needs to be fragmented in order to provide a shield of radius $R_{sh}$ and capable of attenuating the incoming radiation flux to the endurable fraction $\varepsilon$ is given in the first approximation as:

$$R = \sqrt[3]{\frac{3}{4} R_{sh}^2 \frac{(-\ln \varepsilon)}{\mu}}, \qquad (2)$$

(under the assumption of a body in hydrostatic equilibrium, which might not be the case for smaller objects, but still gives us rough measure of the size of required material, and neglecting porosity), $\mu$ is the linear attenuation coefficient, and $\rho$ is the mean density of material. For shielding Earth, we expect

$$R_{sh} = kR_\oplus, \qquad (3)$$

where $k$ is adjustable "safety parameter" of the order unity. If we take, for example, $k$ = 2 (efficiently shielding low Earth orbit and some intermediate orbits), for predominantly icy shielding swarm and $\varepsilon$ = 0.1, we obtain the radius of the parent body as $R(\text{ice}) = 25.4$ km for maximum of emission at 0.1 MeV and $R(\text{ice}) = 34.1$ km (1 MeV). From the point of view of the source located at, for example 10 pc, the target area of the shield spans only about (7.98 $k$) × 10$^{-6}$ arcsec in the sky, making all rays parallel for practical purposes. That radiation from interstellar distances arrives in a parallel beam justifies values like $k$ = 2. For a predominantly rocky body (e.g., an S-type asteroid), the corresponding values are $R(\text{stone}) = 19.1$ km (0.1 MeV) and $R(\text{stone}) = 28.6$ km (1 MeV). Even for a large shield covering geostationary orbit ($k \approx 6.6$), icy shielding would require bodies of effective radius 56.3 km and 75.5 km, respectively. There are many outer Solar System bodies with such modest sizes which could actually provide for the material of the shielding swarm.

Another parameter of interest is the shielding swarm's *porosity*, defined in the standard way (for vacuum- or air-filled systems) as:

$$\phi = 1 - \frac{\rho_{bulk}}{\rho_{part}}, \qquad (4)$$

where $\rho_{bulk}$ and $\rho_{part}$ are the mean densities of the entire swarm and of individual particles comprising the swarm, respectively. For an icy shielding swarm, $\rho_{part} \approx 1$ g cm$^{-3}$, while $\rho_{bulk}$ depends entirely on the engineering solutions for deploying, confining or moving the shielding swarm, so that porosity is effectively a controlling parameter. Arguably, porosity will be significant in any deployed shielding swarm – it is a *swarm*, after all! – and could be changed to avoid second-order effects such as too high scattering rate, or too large reactive forces caused by photoevaporation of particles in the swarm. Since porosity is significant, *bulk density* (the mass of many-particle system of the material divided by the total volume) can be rather small – and it could be changed in order for a particular effect to be produced.

As to the detailed structure and engineering properties of such a shielding swarm we cannot say much at the moment, apart from pointing out in some of the relevant and rapidly developing directions. Ice particle could be controlled by charging them with an ion or electron beam and then pushing with an electric field into a desired configuration. The notion of *smart dust* is particularly interesting in this regard. While attributed to the great SF author and philosopher Stanislaw Lem and his novel *Invincible* [83] by a modern study [84], wide spectrum of applications has become apparent only recently. It has become one of the most discussed concepts in advanced astronautics in the 21st century, e.g. [85-88]. We have no reason to believe that this trend will not continue – to the contrary, flexibility and adaptability of smart dust offers unprecedented prospects for exploratory engineering. In the specific case here, we envision a swarm of particles confined by electromagnetic forces interspersed by smart dust particles controlling the swarm and enabling more precise manipulation, in addition to controlling ionization necessary for the ice particles to be moved around. They could provide essential telemetric information and the data on conditions within the swarm necessary for self-regulation actions. Since various forms of carbon, including fullerenes, is currently thought to be the best material for building smart dust, as well as other nanotechnological applications [89], and the Kuiper Belt objects are carbon-rich, it seems natural to assume that fragmentation of the very same icy body or bodies creating the bulk of the shielding swarm might provide material for construction of smart dust particles as well.

While all this is arguably extremely limited in comparison to what **humanity++** or even just **humanity+** civilization will be able to achieve, the bottom line is that we can see no substantial argument against feasibility of such systems. In contrast – and in the spirit of exploratory engineering – we perceive many possible advantages and merits of such a solution. In simplest term, a possible scenario could be as follows. After a source threatening with cosmic explosion is discovered, long-term monitoring and predictive modelling will be implemented. When the timescale for the explosion is pin-pointed with sufficient precision, the mitigation action plan will be put in motion, by selecting a convenient icy body (and its possible reserves/replenishment sources), changing its orbit in appropriate way to reach the optimal staging grounds. (Of course, if we are dealing with the re-use of already constructed shielding swarm, some of these steps will be skipped.) Upon reaching the staging grounds, the chosen body will be fragmented, and construction of the shielding swarm will begin. Construction phase will be followed by the deployment phase and subsequent mothballing/dispersal of the swarm. In all the manoeuvres, beside the forcing imparted by sources of electromagnetic field, other mechanisms for confinement could be used, notably solar radiation pressure, cf. [90].

While protecting from high-energy photons from a cosmic explosion is the main purpose of mitigation as discussed in this study, one might briefly consider the other possible adverse effects of such explosions. Those discussed so far belong to three categories: (i) neutrino pulse; (ii) cosmic rays; and (iii) direct deposition of matter. As far as neutrinos are concerned, there is not much one could do, at least before **humanity++** level, since no shielding short of neutron-star matter will appreciably attenuate a neutrino beam.[3] Fortunately, early concerns of Collar [92] have been largely showed to be without foundations [93] and the explosion would need to be unrealistically close for neutrinos to produce significant irradiation risk [94].

As mentioned above, whether supernova or GRB explosions present a significant source of astrobiologically important cosmic rays remains controversial [76,95,96]. What is *not* controversial is that cosmic rays have destructive potential from the point of view of biospheres exposed to them. The standard thinking is that GeV cosmic rays, presenting the highest risk, travel ~1 kpc before being significantly deflected by the Galactic magnetic field, so that GRB has to be exceptionally close for its cosmic-ray jet to hit any given planet. Of course, this does not apply to prolonged stay within supernova remnants, or other regions of space with higher-than-average ambiental cosmic ray energy density. While this topic obviously require much further work, it is reasonable to assume that a shielding swarm will cause some absorption and scattering of the cosmic-ray pulse following a very close (and thus undeflected) explosion, water (and water ices) being rather efficient absorber of primary charged particles in the terrestrial conditions. Rough estimates suggest that a shielding swarm with $\varepsilon = 0.1$ (reducing the incoming hard electromagnetic flux tenfold) will, if stabilized over the period of the cosmic ray pulse, reduce fluence of cosmic rays by a factor of a few. In fact, shielding swarms as outlined here, would perhaps be the best solution for cosmic-ray protection as well, since the very same system of electromagnetic confinement necessary for manipulating particles in the swarm could provide additional *active shielding* against incoming charged particles, cf. [97]. While a detailed numerical model is necessary for any conclusion in this respect, it is likely that shielding swarms as envisioned here present the best option for both active and passive shielding of important assets within a single engineering project.

Finally, direct deposition of matter from a cosmic explosion could occur only for very close explosions (on the order of a few pc or less) and hence is quite rare [98]. This process also occurs very slowly, on the timescale of centuries and millennia. While shielding swarms as envisioned will not influence this process much, any civilization on the **humanity+** level will be able to protect itself locally having plenty of time for undertaking any safety measures, if deemed necessary.

---

3   Even **humanity++** civilizations might be hard pressed here, see a fictional example in Egan [91].

Of course, a civilization on the **humanity+** level could construct the shielding closer to Earth, but this seems to be less convenient for several reasons. Not only is the adequate material less available and minor gain in size achieved at a farther out location, but it is also more likely to be already harvested for industrial purposes by the time the mitigation project becomes actual. Outer Solar System provides low temperature at which ice is by far the most convenient building material, especially in its high-density phase which occurs only at very low temperatures, below about 30 K [99]. Inner Solar system is likely to be filled with pieces of the "technosphere" – and construction of a planet-size swarm of objects , even if of low total mass, in such environment would introduce an unnecessary debris collision risk. A further advantage of using a swarm of objects rather than a single solid object or a small number of solid objects is that while the collision risk stays the same (or is probably even slightly enhanced for the swarm, due to incomplete containment) as long as the cross-section is the same, the *effects* of a collision might be much less for a swarm. Indeed, controlled, slow collisions might be desirable as the source of further material for replenishing the swarm, which will be occasionally necessary (see below).

The risk will be increased by scattered γ-radiation and possible secondary cosmic rays created by the shielding – so this is another reason to keep the shield as distant as possible. Moreover, since the shielding swarm could be mothballed for possible future use, the efficiency of the resource utilization will generally increase. Once developed, the technology could then be used in multiple contexts, including those scaled-down for local safety purposes. In the unlikely case that **humanity+** civilization has not renounced aggressive and warlike tendencies, shielding swarms could be used for defence of habitats and installations against long-range radiation and particle weaponry as well.

An important counterargument relates to the required bulk motion of the shielding swarm.[4] Anything put closer to the Sun will need less active (non-Keplerian) motion to retain favourable alignment with Earth over time. Since the uncertainty over the exact timing of the explosion translates into capability of maintaining the exact alignment over comparable period of time – including a liberal safety margin – more distant shielding will require higher energy expenditure (more rocket fuel, effectively) than a closer one for the same performance parameters. In other words, larger acceleration for the same period of time will be required for distant shielding than for a closer one. A more detailed quantitative analysis is needed to establish how this effect impacts the overall economic side of the story. Relatively smaller mass and higher mobility of the shielding swarms will still maintain a practical advantage over any kind of solid shields of comparable efficiency.

Minor shields, of course, could be built along similar lines, for objects and installations deemed particularly vulnerable to the effects of cosmic explosions. For example, generic shielding of O'Neill habitats would perhaps be found too light to withstand the blast of electromagnetic/cosmic-ray radiation following a close supernova or a GRB, and additional shielding required. Of course, seemingly more practical course of action could be to bring the habitats into the shadow of a large planet, like Jupiter or Neptune. However, building of a shielding swarm might still be better option, particularly if a habitat is in a remote heliocentric orbit either very close to the Sun or very far away (in the Kuiper belt or farther) and there is a necessity of keeping it in a shadow for a prolonged period of time. In the latter case, defying orbital motion could cost a significant amount of energy, possibly much larger than assembling and operating the shielding swarm. The advantage of smaller shielding swarms could as well be in the operational timescale as well: while the swarm of kilometer- or tens of kilometers-scale could be assembled on the scale of years even if we assume "snail pace" characteristic velocities of $10^{-3}$ m s$^{-1}$, realistic timescales for moving the correspondingly large habitat from another part of the Solar system to Jupiter's or Neptune's shadow could take much longer under inertial motion.

---

[4] The authors thank an anonymous referee for bringing attention to this important point.

A shielding swarm will be subject to wear and tear in the course of its utilization for at least two reasons: (i) ablation from both constant and intermittent radiation exposure; and (ii) loss of components due to imperfect confinement. Both of these are easily modelled, predicted and countered by inserting extra components from a surrounding or close reservoir. (This is another reason why shielding swarms are better constructed in the outer part of a planetary system.) (ii) can be generalized to encompass effects of diffusion, undesirable fragmentation and/or adhesion and similar processes. Evolution of such a complex system is already tractable by known methods of statistical and condensed-matter physics; there is no reason to doubt that our understanding will not only improve in future on timescales much smaller than those for transition to **humanity++** state.

## 4. A new astroengineering signature

In the last decade or so, the interest in finding astro-engineering signatures of ATCs has flared up again, motivated by both observational and theoretical insights [100-105]. In addition, considerations related to the future of humanity and awareness of the possibilities of future (post)human astroengineering, e.g. [64], have led to proposals for many such projects and better understanding of the issues involved.

This is an extension of the Dysonian "mirroring" – anything which we can hypothesize about humanity's future should apply to at least some extraterrestrial intelligent species as well [61,106]. If we, for example, eventually become capable and willing to build Dyson shells around Sun and possibly other nearby stars, it makes sense to search for Dyson shells around other stars. Of course, this pertains to those astroengineering projects having clear and unequivocal utility stemming from general considerations of physics and economics (and not, for example, to those constructed for artistic purposes, although such are possible and, in the fullness of time, even probable). From the point of view of both the engineers of the future humanity and of any other extraterrestrial intelligent species, there can be few larger utilities and hence stronger motivations for undertaking such huge enterprises than gaining resistance, if not immunity, to large catastrophic risks stemming from cosmic explosions. Even if those are not truly existential risks at that level of civilizational development, the negative utility of an adverse outcome (unmitigated explosion of close supernova or GRB or a magnetar) can be so great to trump any other considerations, especially those pertaining to the price of such mitigation endeavour. One might say that the reasoning of Bostrom [4] about existential risks threatening present day/near future humanity generalizes adequately to **humanity+** and **humanity++** – and their extraterrestrial intelligent analogs. Since this type of artefact pertains to reduction in large risk, it is rational *both* to build such artefacts (for civilizations which have capabilities) and to search for such artefacts (for civilizations interested in SETI-analogous activities).

So there is much more at stake in attempting to detect traces and manifestations of astroengineering from afar. Of course, it is very difficult (at least), to give confident assertions about the future astroengineering capacities – which translates into uncertainty about their detection signatures. Above, we have proposed than a novel form of astroengineering undertaken by advanced technological civilizations in the Galaxy (but perhaps not superadvanced like **humanity++**) consists in building shielding swarms, probably in the outer parts of their planetary systems. Since the whole process needs to involve lots of technological activity *in situ*, some astroengineering signatures could be detected, especially if located far from the glare of the parent star and in regions of low ambiental temperature, where infrared signatures are easiest to detect [103,104,107]. While much further work on the *exploratory engineering* (cf. Reference 108) is necessary to ascertain the details, it is at least *plausible* to expect some or all of the following items could be observed:

> ➢ Structures of planetary size (terrestrial planets and larger) with anomalously small masses and non-Keplerian motion. While it is difficult to establish masses of objects without tacitly assuming their naturalistic origin – the "catch-22" of exploratory engineering – observing very low masses $\sim 10^{-8} M_\oplus$ for Earth-sized objects would be a strong indication that we are looking at an artefact.
> ➢ Anomalous optical properties such as polarization and non-equilibrium temperatures are further detectable signs. If the swarming shield is predominantly made of ice, we should expect very strong absorption in the far-infrared around 1–10 μm, as well as presence of water vapour and OH absorption bands from an extended envelope. If the shield is transiting, one should expect small optical transit depths coupled with much larger infrared transit depths.
> ➢ Fragmentation of small bodies in a planetary system without obvious physical causes such as collisions would indicate astroengineering. This might be accompanied by cascade fragmentation of debris and apparently anomalous (from the point of view of a distant observer) loss of kinetic energy and momentum.
> ➢ Temporal and spatial coincidences between the last phases of evolution of supernova progenitors, as well as magnetar explosions, and anomalous activities in a planetary system containing one or more habitable planets.[5] In the case of a supernova, only its immediate Galactic environment up to a few of tens of parsecs should be monitored for such coincidences.

## 5. Conclusions

We have considered possible long-term strategies future humans and/or other advanced Galactic civilizations could employ in order to mitigate the threat of supernova/GRB/ magnetar explosions in their astronomical vicinity. While the destructive potential of such dramatic events has never been in doubt, their low frequency and multiple complex ecological effects have made them unappealing subjects for the risk analysis/mitigation studies until very recently.

---

[5] GRBs are hardly relevant here since they tend to influence the entire Galactic Habitable Zone. On the other hand, the probability of a Galactic GRB occurring while humanity is interested in SETI activities is clearly minuscule.

We have emphasized the predictability of cosmic explosions as the key factor in any mitigation strategy. Astrophysical processes leading to those explosions are not entirely clear so far, but the progress in both observational and theoretical astrophysics justifies the conclusion that this is the easiest of the conditions to be satisfied, unless some catastrophe of different kind arrests/stops the progress of science in the near future. Other pre-conditions dealing with industrial base of advanced civilizations in the outer parts of their planetary systems are common to many future-studies scenarios.

If it turns out that a cosmic explosion is imminent, mitigation will be undertaken. While mitigation might take many forms, we have argued above that the most reasonable is local construction of shielding swarms. Shielding swarms could be made of many materials available to an advanced technological society, but one of the simplest options would be ice particles, with an admixture of smart dust as part of the confining mechanism. Such systems possess a number of desirable properties and are not entirely unrealistic from an engineering perspective (i.e., there are no unresolvable *systemic* obstacles for their construction, in contrast to warp drives, faster-than-light travel, closed timelike curves, solid Dyson shells, and the like). Most importantly, shielding swarms are *inexpensive* to build, move, deploy, utilize, preserve, and decommission. Future precise, quantitative work – highly desirable as in other cases of exploratory engineering – is not likely to change that desirable property. Their efficiency in mitigating γ-rays from cosmic explosion is rather high, although the issue of high-energy cosmic ray jets of the same sources remains unclear and potentially troubling.

In brief, we conclude that successful mitigation of cosmic explosions risk is viable for sufficiently advanced technological societies, both future terrestrial and extraterrestrial. We suggest that building and maintaining shielding swarms of small particles/components is (relatively!) cheap and efficient way of achieving that goal and creating a durable planetary and interplanetary civilization. The technology required partially overlaps with that required for mitigation of asteroid/cometary impact risk, which could provide some clues for future technological desiderata and even convergence. Finally, this new type of macro- or astroengineering could not only enrich the spectrum of astroengineering possibilities, but also provide another opportunity for bold and innovative SETI programs to detect advanced technological civilizations elsewhere in the Galaxy.

**Acknowledgements**. The authors wish to express gratitude to two anonymous referees for comments and suggestions which improved the quality of a previous versions of the manuscript. M.M.Ć. thanks Jelena Dimitrijević, Nick Bostrom, George Dvorsky, Paul Gilster, Vojin Rakić, and Slobodan Popović for many useful and pleasant discussions of the future-related topics as well as kind encouragement during the work on this project. B. V. thanks Dominik Steinhauser for great help with N-body simulations. The authors acknowledge financial support from the Ministry of Education, Science and Technological Development of the Republic of Serbia through the project #ON176021 'Visible and invisible matter in nearby galaxies: theory and observations'.